\documentclass{llncs}
\usepackage{makeidx}
\usepackage{graphicx}

\begin{document}

%\pagestyle{headings}  % switches on printing of running heads
%\addtocmark{XWeB: the XML Warehouse Benchmark} % additional mark in the TOC

\mainmatter 

\title{XWeB: the XML Warehouse Benchmark}
\titlerunning{XWeB: the XML Warehouse Benchmark}
\toctitle{XWeB: the XML Warehouse Benchmark}

\author{Hadj Mahboubi \inst{1} \and J\'{e}r\^{o}me Darmont \inst{2}}
\authorrunning{Hadj Mahboubi and J\'{e}r\^{o}me Darmont}   % abbreviated author list (for running head)
\tocauthor{Hadj Mahboubi, J\'{e}r\^{o}me Darmont}

\institute{CEMAGREF Clermont-Ferrand\\
24 avenue des Landais, BP 50085, 63172 Aubi\`{e}re Cedex, France\\
\email{hadj.mahboubi@cemagref.fr} -- \texttt{http://eric.univ-lyon2.fr/\homedir hmahboubi/}
\and
Universit\'{e} de Lyon (ERIC Lyon 2)\\
5 avenue Pierre Mend\`{e}s-France, 69676 Bron Cedex, France\\
\email{jerome.darmont@univ-lyon2.fr} -- \texttt{http://eric.univ-lyon2.fr/\homedir jdarmont/}
}

\maketitle

\begin{abstract}
With the emergence of XML as a standard for representing business data, new decision support applications are being developed. These XML data warehouses aim at supporting On-Line Analytical Processing (OLAP) operations that manipulate irregular XML data. To ensure feasibility of these new tools, important performance issues must be addressed. Performance is customarily assessed with the help of benchmarks. However, decision support benchmarks do not currently support XML features.
In this paper, we introduce the XML Warehouse Benchmark (XWeB), which aims at filling this gap. XWeB derives from the relational decision support benchmark TPC-H. It is mainly composed of a test data warehouse that is based on a unified reference model for XML warehouses and that features XML-specific structures, and its associate XQuery decision support workload. XWeB's usage is illustrated by experiments on several XML database management systems.
\keywords{benchmark, XML data warehouse, OLAP, TPC-H}
\end{abstract}

\section{Introduction}
\label{sec:Introduction}

With the increasing volume of XML data available, and XML now being a standard for representing complex business data~\cite{beyer05}, XML data sources that are pertinent for decision support are ever more numerous. However, XML data bear irregular structures (e.g., optional and/or diversely ordered elements, ragged hierarchies, etc.) that would be intricate to handle in a relational Database Management System (DBMS). Therefore, many efforts toward XML data warehousing have been achieved~\cite{dwdaea09,P02,ZWLZ05}, as well as efforts for extending the XQuery language with On-Line Analytical Processing (OLAP) capabilities~\cite{datax08,KitAK09,WiwatwattanaJLS07}.

XML-native DBMSs supporting XQuery should naturally form the basic storage component of XML warehouses. However, they currently present relatively poor performances when dealing with the large data volumes and complex analytical queries that are typical in data warehouses, and are thus challenged by relational, XML-compatible DBMSs. A tremendous amount of research is currently in progress to help them become a credible alternative, though. Since performance is a critical issue in this context, its assessment is primordial. 

Database performance is customarily evaluated experimentally with the help of benchmarks. However, existing decision support benchmarks~\cite{ijbidm06,APB1,oneil09,TCP05c} do not support XML features, while XML benchmarks~\cite{BohmeR02,BressanLLLN02,SWKCMB02,YaoOK04} target transactional applications and are ill-suited to evaluate the performances of decision-oriented applications. Their database schemas do not bear the multidimensional structure that is typical in data warehouses (i.e., star schemas and derivatives bearing facts described by dimensions~\cite{kimball}); and their workloads do not feature typical, OLAP-like analytic queries. 

Therefore, we present in this paper the first (to the best of our knowledge) XML decision support benchmark. Our objective is to propose a test XML data warehouse and its associate XQuery decision support workload, for performance evaluation purposes. The XML Warehouse Benchmark (XWeB) is based on a unified reference model for XML data warehouses~\cite{dwdaea09}. An early version of XWeB~\cite{asd06} was derived from the standard relational decision support benchmark TPC-H~\cite{TPCH}. In addition, XWeB's warehouse model has now been complemented with XML-specific irregular structures, and its workload has been both adapted in consequence and expanded.

The remainder of this paper is organized as follows. In Section~\ref{sec:RelatedWork}, we present and discuss related work regarding relational decision support and XML benchmarks. In Section~\ref{sec:ReferenceXMLWarehouseModel}, we recall the XML data warehouse model XWeB is based on. In Section~\ref{sec:XWeBSpecifications}, we provide the full specifications of XWeB. In Section~\ref{sec:Experiments}, we illustrate our benchmark's usage by experimenting on several XML DBMSs. We finally conclude this paper and provide future research directions in Section~\ref{sec:Conclusion}.

\section{Related Work}
\label{sec:RelatedWork}

\subsection{Relational Decision Support Benchmarks}
\label{rdsb}

The OLAP APB-1 benchmark has been very popular in the late nineties~\cite{APB1}. Issued by the OLAP Council, a now inactive organization founded by four OLAP solution vendors, APB-1's data warehouse schema is structured around \emph{Sale} facts and four dimensions: \emph{Customer}, \emph{Product}, \emph{Channel} and \emph{Time}. Its workload of ten queries aims at sale forecasting. Although APB-1 is simple to understand and use, it proves limited, since it is not ``differentiated to reflect the hurdles that are specific to different industries and functions"~\cite{T98}.

Henceforth, the Transaction Processing Performance Council (TPC) defines standard benchmarks and publishes objective and verifiable performance evaluations to the industry. The TPC currently supports one decision support benchmark: TPC-H~\cite{TPCH}. TPC-H's database is a classical \emph{product-order-supplier} model. Its workload is constituted of twenty-two SQL-92, parameterized, decision support queries and two refreshing functions that insert tuples into and delete tuples from the database, respectively. Query parameters are randomly instantiated following a uniform law. Three primary metrics are used in TPC-H. They describe performance in terms of power, throughput, and a combination of these two criteria. Power and throughput are the geometric and arithmetic mean values of database size divided by workload execution time, respectively.

Although decision-oriented, TPC-H's database schema is not a typical star-like data warehouse schema. Moreover, its workload does not include any explicit OLAP query. The TPC-DS benchmark, which is currently in its latest stages of development, fixes this up~\cite{TCP05c}. TPC-DS' schema represents the decision support functions of a retailer under the form of a constellation schema with several fact tables and shared dimensions. TPC-DS' workload is constituted of four classes of queries: reporting queries, ad-hoc decision support queries, interactive OLAP queries, and extraction queries. SQL-99 query templates help randomly generate a set of about five hundred queries, following non-uniform distributions. The warehouse maintenance process includes a full Extract, Transform and Load (ETL) phase, and handles dimensions with respect to their nature (non-static dimensions scale up while static dimensions are updated). One primary throughput metric is proposed in TPC-DS to take both query execution and the maintenance phase into account.

More recently, the Star Schema Benchmark (SSB) has been proposed as a simpler alternative to TPC-DS \cite{oneil09}. As our early version of XWeB~\cite{asd06}, it is based on TPC-H's database remodeled as a star schema. It is basically architectured around an \emph{order} fact table merged from two TPC-H tables. But more interestingly, SSB features a query workload that provides both functional and selectivity coverages. 

As in all TPC benchmarks, scaling in TPC-H, TPC-DS and SSB is achieved through a scale factor $SF$ that helps define database size (from 1~GB to 100~TB). Both database schema and workload are fixed. The number of generated queries in TPC-DS also directly depends on $SF$.
TPC standard benchmarks aim at comparing the performances of different systems in the same experimental conditions, and are intentionally not very tunable.  By contrast, the Data Warehouse Engineering Benchmark (DWEB) helps generate various ad-hoc synthetic data warehouses (modeled as star, snowflake, or constellation schemas) and workloads that include typical OLAP queries~\cite{ijbidm06}. DWEB targets data warehouse designers and allows testing the effect of design choices or optimization techniques in various experimental conditions. Thus, it may be viewed more like a benchmark generator than an actual, single benchmark. DWEB's main drawback is that its complete set of parameters makes it somewhat difficult to master.

Finally, to be complete, TPC-H and TPC-DS have recently be judged insufficient for ETL purposes \cite{simitsis09} and specific benchmarks for ETL workflows are announced \cite{simitsis09,wyatt09}.

\subsection{XML Benchmarks}

XML benchmarks may be subdivided into two families. On one hand, micro-benchmarks, such as the Michigan Benchmark (so-named in reference to the relational Wisconsin Benchmark developed in the eighties) \cite{RPJCA06} and MemBeR~\cite{AMM05}, help XML documents storage solution designers isolate critical issues to optimize. More precisely, micro-benchmarks aim at assessing the individual performances of basic operations such as projection, selection, join and aggregation. These low-level benchmarks are obviously too specialized for decision support application evaluation, which requires testing complex queries at a more global level.

On the other hand, application benchmarks help users compare the global performances of XML-native or compatible DBMSs, and more particularly of their query processor. For instance, X-Mach1~\cite{BohmeR02}, XMark~\cite{SWKCMB02}, XOO7 (an extension of the object-oriented benchmark OO7) \cite{BressanLLLN02} and XBench~\cite{YaoOK04} are application benchmarks. Each implements a mixed XML database that is both data-oriented (structured data) and document-oriented (in general, random texts built from a dictionary). However, except for XBench that proposes a true mixed database, their orientation is either more particularly focused on data (XMark, XOO7) or documents (X-Mach1). 

These benchmarks also differ in: the fixed or flexible nature of the XML schema (one or several Document Type Definitions -- DTDs -- or XML Schemas); the number of XML documents used to model the database at the physical level (one or several); the inclusion or not of update operations in the workload. We can also underline that only XBench helps evaluate all the functionalities offered by the XQuery language. Unfortunately, none of these benchmarks exhibit any decision support feature. This is why relational benchmarks presented in Section~\ref{rdsb} are more useful to us in a first step. 

\section{Reference XML Warehouse Model}
\label{sec:ReferenceXMLWarehouseModel}

Existing XML data warehouse architectures more or less converge toward a unified model. They mostly differ in the way dimensions are handled and the number of XML documents that are used to store facts and dimensions. Searching for the best compromise in terms of query performance and modeling power, we proposed a unified model \cite{dwdaea09} that we reuse in XWeB. As XCube \cite{HBH03}, our reference XML warehouse is composed of three types of XML documents at the physical level:
	 document \emph{dw-model.xml} defines the multidimensional structure of the warehouse (metadata);
	 each $facts_f.xml$ document stores information related to set of facts $f$ (several fact documents allow constellation schemas);
	 each $dimension_d.xml$ document stores a given dimension $d$'s member values for any hierarchical level.	

\begin{figure}[hbt]
{\centering
\resizebox*{0.8\textwidth}{!}{\includegraphics{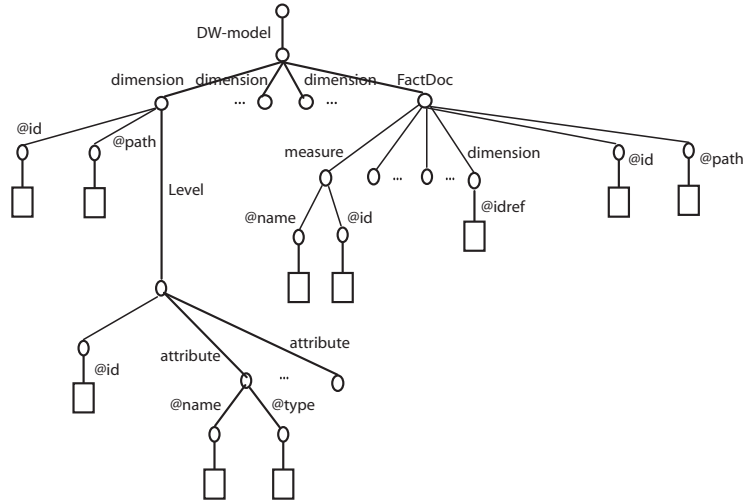}}
\par}
\caption{\textit{dw-model.xml} graph structure} \label{fig:dw-medel}
\end{figure}

\begin{figure*}[hbt]
{\centering
\resizebox*{1.0\textwidth}{!}{\includegraphics{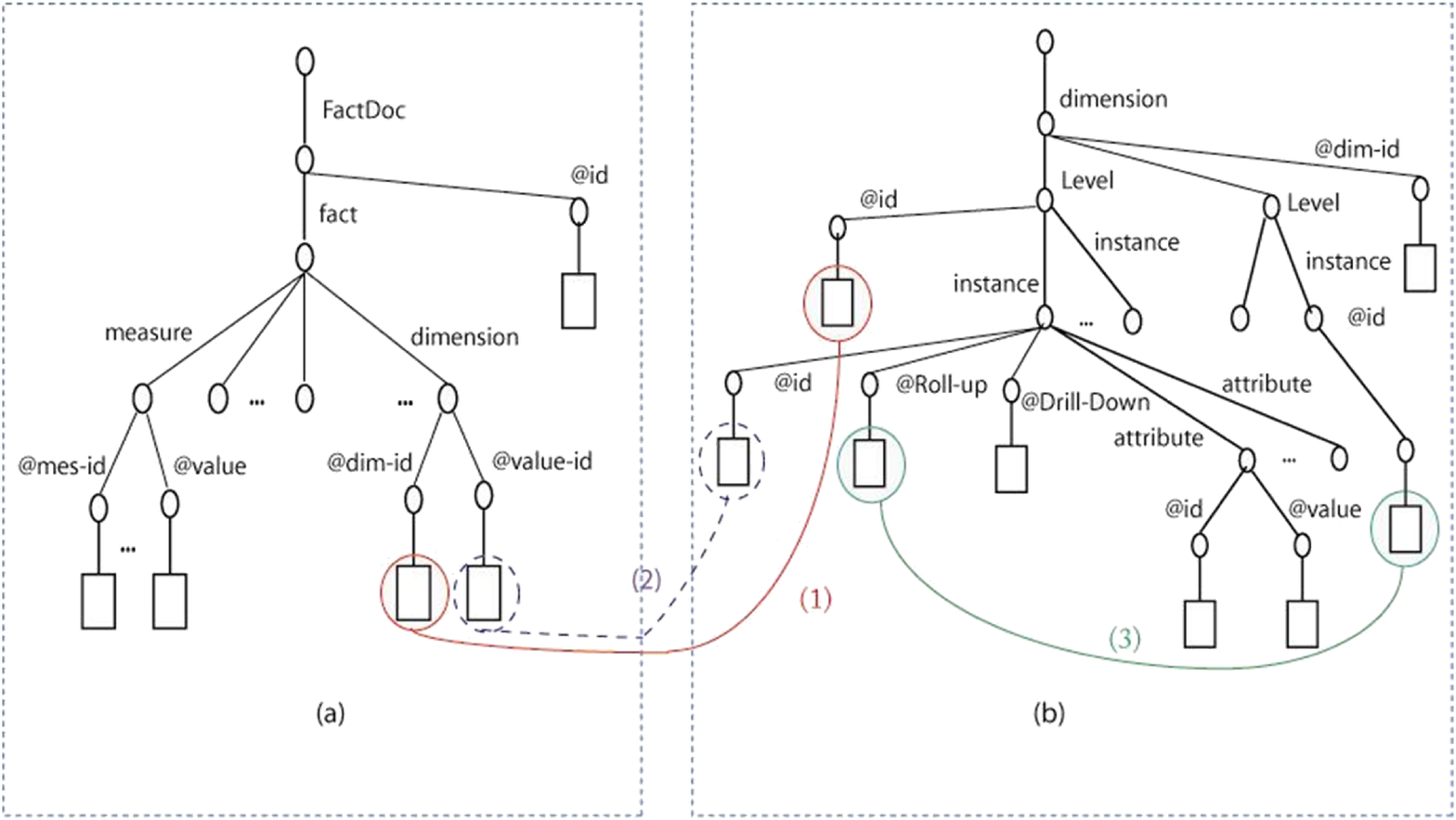}}
\par}
\caption{$facts_f.xml$ (a) and $dimension_d.xml$ (b) graph structures} \label{fig:fact-dimension}
\end{figure*}

More precisely, \textit{dw-model.xml}'s structure (Figure~\ref{fig:dw-medel}) bears two types of nodes: $dimension$ and $FactDoc$ nodes. A $dimension$ node defines one dimension, its possible hierarchical levels ($Level$ elements) and attributes (including types), as well as the path to the corresponding $dimension_d.xml$ document. A $FactDoc$ element defines a fact, i.e., its measures, references to the corresponding dimensions, and the path to the corresponding $facts_f.xml$ document.
The $facts_f.xml$ documents' structure (Figure~\ref{fig:fact-dimension}(a)) is composed of $fact$ subelements that each instantiate a fact, i.e., measure values and dimension references. These identifier-based references support the fact-to-dimension relationships.

Finally, the $dimension_d.xml$ documents' structure (Figure~\ref{fig:fact-dimension}(b)) is composed of \textit{Level} nodes. Each of them defines a hierarchy level composed of \textit{instance} nodes. An \textit{instance} defines the member attributes of a hierarchy level as well as their values.

\section{XWeB Specifications}
\label{sec:XWeBSpecifications}

\subsection{Principle}
\label{sec:Principle}

XWeB derives from TPC-H, modified in a number of ways explained in the following sections, for three reasons. First, we acknowledge the importance of TPC benchmarks' standard status. Hence, our goal is to have XWeB inherit from TPC-H's wide acceptance and usage (whereas TPC-DS is still under development). Second, from our experience in designing the DWEB relational data warehouse benchmark, we learned that Gray's simplicity criterion for a good benchmark \cite{GRA93} is primordial. This is again why we preferred TPC-H, which is much simpler than TPC-DS or DWEB. Third, from a sheer practical point of view, we also selected TPC-H to benefit from its data generator, \texttt{dbgen}, a feature that does not exist in TPC-DS yet.

The main components in a benchmark are its database and workload models. XWeB's are described in Sections~\ref{sec:Database} and \ref{sec:Workload}, respectively. In a first step, we do not propose to include ETL features in XWeB, although XQuery has been complemented with update queries recently \cite{xqu09}. ETL is indeed a complex process that presumably requires dedicated benchmarks 
\cite{simitsis09}. Moreover, the following specifications already provide a raw loading evaluation framework. The XWeB warehouse is indeed a set of XML documents that must be loaded into an XML DBMS, an operation that can be timed.

\subsection{Database Model}
\label{sec:Database}

\subsubsection{Schema.}

\begin{figure}[hbt]
{\centering
\resizebox*{1.0\textwidth}{!}{\includegraphics{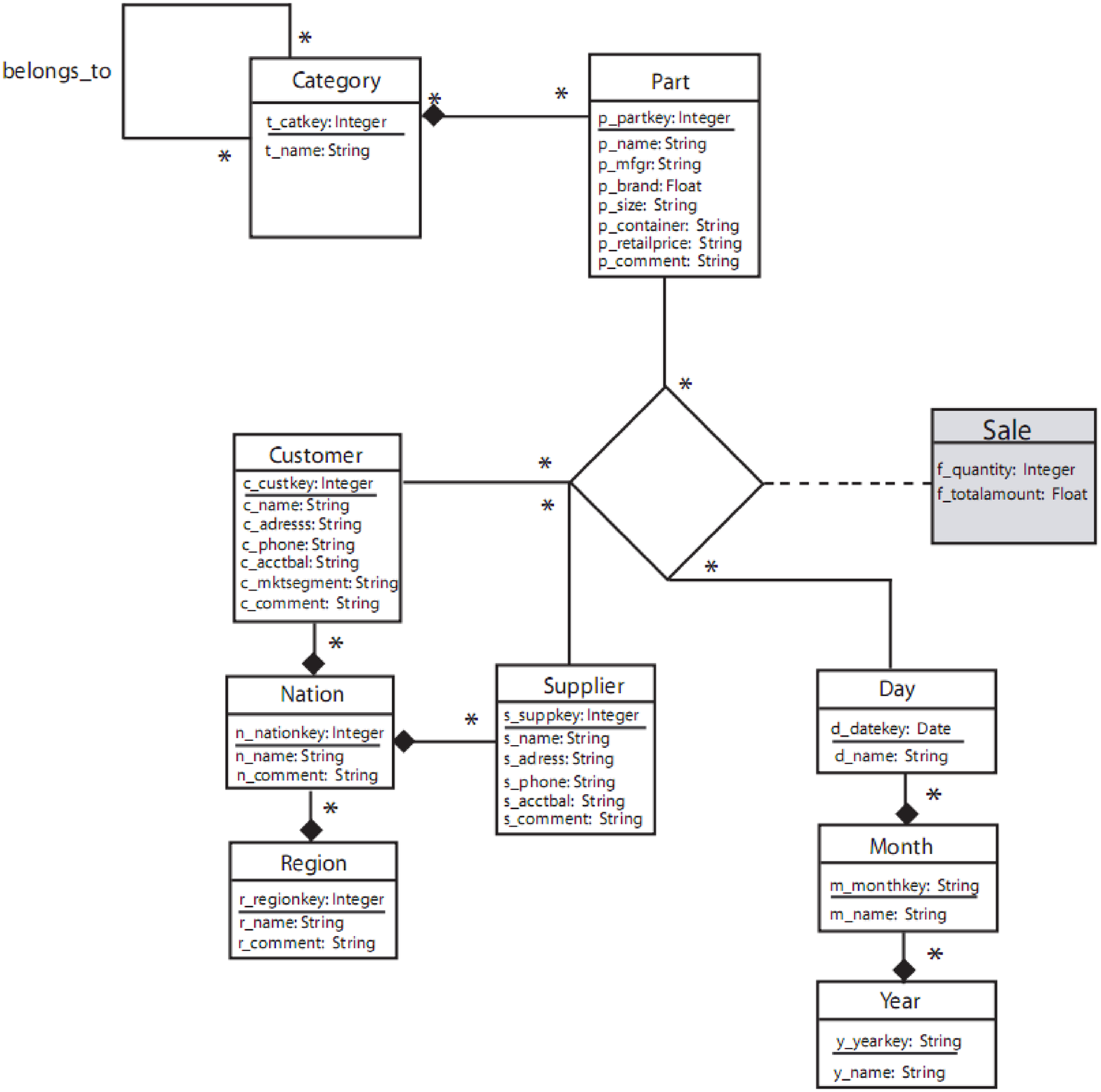}}
\par}
\caption{XWeB warehouse's conceptual schema} \label{fig:schema}
\end{figure}

At the conceptual level, like O'Neil et al. in SSB, we remodel TPC-H's database schema as an explicit multidimensional (snowflake) schema (Figure~\ref{fig:schema}), where \emph{Sale} facts are described by the \emph{Part/Category}, \emph{Customer/Nation/Region}, \emph{Supplier/Nation/Region} and \emph{Day/Month/Year} dimensions. 

The \emph{Part/Category} hierarchy, which is not present in TPC-H, is of particular interest. It is indeed both non-strict and non-covering \cite{mdb03t}. Beyer et al. would term it \emph{ragged} \cite{beyer05}. We prefer the term \emph{complex} since ragged hierarchy has different meanings in the literature; e.g., Rizzi defines it as non-covering only \cite{dwo07r}. More precisely, in our context, non-strictness means relationships between parts and categories, and between categories themselves, are many-to-many.  Non-coveringness means parts and subcategories may roll up to categories at any higher granularity level, i.e., skipping one or more intermediary granularity levels. Complex hierarchies do exist in the real world, are easy to implement in XML, whereas they would be intricate to handle in a relational system \cite{beyer05}.

At the logical level, the UML class diagram from Figure~\ref{fig:schema} translates into an instance of \emph{dw-model.xml} (Figure~\ref{fig:xmlschema}). Attributes (fact measures and dimension members) are not mentioned in Figure~\ref{fig:xmlschema} for brevity, but they are present in the actual document.

\begin{figure}[hbt]
\begin{verbatim}
<?xml version="1.0" encoding="UTF-8"?>
<xweb-dw-model>
  <fact id="Sale" path="f_sale.xml"/>
  <dimension id="Date" path="d_date.xml">
    <level id="Day" rollup="Month" drilldown=""/>
    <level id="Month" rollup="Year" drilldown="Day"/>
    <level id="Year" rollup="" drilldown="Month"/>
  </dimension>
  <dimension id="PartDim" path="d_part.xml"/>
    <level id="Part" rollup="Category" drilldown=""/>
    <level id="Category" rollup="Category" drilldown="Part Category"/>
  </dimension>
  <dimension id="CustomerDim" path="d_customer.xml">
    <level id="Customer" rollup="C_Nation" drilldown=""/>
    <level id="C_Nation" rollup="C_Region" drilldown="Customer"/>
    <level id="C_Region" rollup="" drilldown="C_Nation"/>
  </dimension>
  <dimension id="SupplierDim" path="d_supplier.xml">
    <level id="Supplier" rollup="S_Nation" drilldown=""/>
    <level id="S_Nation" rollup="S_Region" drilldown="Supplier"/>
    <level id="S_Region" rollup="" drilldown="S_Nation"/>
  </dimension>
</xweb-dw-model>
\end{verbatim}
\caption{XWeB warehouse's logical schema} 
\label{fig:xmlschema}
\end{figure}

Finally, at the physical level, fact and dimension instances are stored in a set of XML documents, namely $facts_1.xml = f\_sale.xml$, $dimension_1 = d\_date.xml$, $dimension_2 = d\_part.xml$, $dimension_3 = d\_customer.xml$ and $dimension_4 = d\_supplier.xml$. To introduce further XML-specific features in XWeB, $f\_sale.xml$'s DTD allows missing dimension references and measures, as well as any order in fact subelements. Our aim here is to introduce a measure of ``dirty data'' in the benchmark.

\subsubsection{Parameterization.}
\label{param}

XWeB's main parameters basically help control data warehouse size. Size ($S$) depends on two parameters: the scale factor ($SF$) inherited from TPC-H, and density $D$. When $D = 1$, all possible combinations of dimension references are present in the fact document (Cartesian product), which is very rare in real-life data warehouses. When $D$ decreases, we progressively eliminate some of these combinations. $D$ actually helps control the overall size of facts independently from the size of dimensions.

$S$ can be estimated as follows: $S = S_{dimensions} + S_{facts}$, where $S_{dimensions}$ is the size of dimensions, which does not change when $SF$ is fixed, and $S_{facts}$ is the size of facts, which depends on $D$. \begin{math} S_{dimensions} = \sum_{d \in \mathcal D} |d|_{SF} \times nodesize(d)\end{math} and  \begin{math} S_{facts} = \prod_{d \in \mathcal D} |h^d_1|_{SF} \times D \times fact\_size \end{math}, where $\mathcal D$ is the set of dimensions, $|d|_{SF}$ the total size of dimension $d$ (i.e., all hierarchy levels included) w.r.t. $SF$, $|h^d_1|_{SF}$ the size of the coarsest hierarchy level in dimension $d$ w.r.t. $SF$, $nodesize(d)$ the average node size in $dimension_{d}.xml$, and $fact\_size$ the average fact element size. For example, when $SF = 1$ and $D = 1$, with node sizes all equal to 220 bytes, the size of $f\_sale.xml$ is 2065~GB.
Eventually, two additional parameters control the probability of missing values ($P_m$) and element reordering ($P_o$) in facts, respectively.

\subsubsection{Schema Instantiation.}

The schema instantiation process is achieved in two steps: first, we build dimension XML documents, and then the fact document. 
Dimension data are obtained from \texttt{dbgen} as flat files. Their size is tuned by $SF$. Dimension data are then matched to the \textit{dw-model.xml} document, which contains dimension specifications, hierarchical levels and attribute names, to output the set of $dimension_d.xml$ ($d \in \mathcal D$) documents. $d\_part.xml$, which features a complex hierarchy, is a particular case that we focus on. 

Algorithm from Figure~\ref{cats-algo} describes how categories are assigned to parts from $d\_part.xml$. First, category names are taken from TPC-H and organized in three arbitrary levels in the $cat$ table. Moreover, categories are interrelated through rollup and drill-down relationships to form a non-strict hierarchy. For example, level-2 category BRUSHED rolls up to level-1 categories NICKEL and STEEL, and drills down to level-3 categories ECONOMY, STANDARD and SMALL. The whole hierarchy extension is available on-line (Section~\ref{sec:Conclusion}). Then, to achieve non-coveringness, we assign to each part $p$ several categories at any level. $p.catset$ denotes the set of categories assigned to $p$. Each ``root'' category (numbering from 1 to 3) is selected from a random level $lvl$. Then, subcategories may be (randomly) selected from subsequent levels. Non-coveringness is achieved when initial level is lower than 3 and there is no subcategory. $ncat$ and $nsubcat$ refer to category and subcategory numbers, respectively. $cand$ denotes a candidate category or subcategory. $|cat[i]|$ is the number of elements in table $cat$'s $i^{th}$ level.

\begin{figure}[hbt]
\begin{verbatim}
cat := [[BRASS, COPPER, NICKEL, STEEL, TIN],                // level 1
        [ANODIZED, BRUSHED, BURNISHED, PLATED, POLISHED],   // level 2
        [ECONOMY, LARGE, MEDIUM, PROMO, SMALL, STANDARD]]   // level 3
FOR ALL p IN d_part DO
   p.catset := EMPTY_SET
   ncat := RANDOM(1, 3)
   FOR i := 1 TO ncat DO
      lvl := RANDOM(1, 3)
      REPEAT 
         cand := cat[lvl, RANDOM(1, |cat[lvl]|)]
      UNTIL cand NOT IN p.catset
      p.catset := p.catset UNION cand
      nsubcat := RANDOM(0, 3 - lvl)
      FOR j := 1 TO nsubcat DO
         cand := cat[lvl + j, RANDOM(1, |cat[lvl + j]|)]
         IF cand NOT IN p.catset THEN
            p.catset := p.catset UNION cand
         END IF
      END FOR
   END FOR
END FOR   
\end{verbatim}
\caption{Part category selection algorithm}
\label{cats-algo}
\end{figure}

Facts are generated randomly with respect to the algorithm from Figure~\ref{facts-algo}. The number of facts depends on $D$, and data dirtiness on $P_m$ and $P_o$ (Section~\ref{param}). $D$, $P_m$ and $P_o$ are actually used as Bernouilli parameters. $val$ is a transient table that stores dimension references and measure values, to allow them to be nullified and/or reordered without altering loop index values. The \texttt{SKEWED\_RANDOM()} function helps generate ``hot'' and ``cold'' values for measures $Quantity$ and $TotalAmount$, which influences range queries. Finally, the \texttt{SWITCH()} function randomly reorders a set of values. 

\begin{figure}[hbt]
\begin{verbatim}
FOR ALL c IN d_customer DO
   FOR ALL p IN d_part DO
      FOR ALL s IN d_supplier DO
         FOR ALL d IN d_date DO
            IF RANDOM(0, 1) <= D THEN
               // Measure random generation
               Quantity := SKEWED_RANDOM(1, 10000)  
               TotalAmount := Quantity * p.p_retailprice
               // Missing values management
               val[1] := c.c_custkey; val[2] := p.p_partkey
               val[3] := s.s_suppkey; val[4] := d.d_datekey
               val[5] := Quantity; val[6] := TotalAmount
               FOR i := 1 TO 6 DO
                  IF RANDOM(0, 1) <= Pm THEN
                     val[i] := NULL
                  END IF
               END FOR
               // Dimension reordering
               IF RANDOM(0, 1) <= Po THEN
                  SWITCH(val)
               END IF
               WRITE(val) // Append current fact into f_sale.xml
            END IF
         END FOR
      END FOR
   END FOR
END FOR
\end{verbatim}
\caption{Fact generation algorithm}
\label{facts-algo}
\end{figure}

\subsection{Workload Model}
\label{sec:Workload}

\subsubsection{Workload Queries and Parameterization.}
\label{sec:WorkloadQueries}

The XQuery language \cite{XQ04} allows formulating decision support queries, unlike simpler languages such as XPath. Complex queries, including aggregation operations and join queries over multiple documents, can indeed be expressed with the FLWOR syntax. However, we are aware that some analytic queries are difficult to express and execute efficiently with XQuery, which does not include an explicit grouping construct comparable to the \texttt{GROUP BY} clause in SQL \cite{beyer05}. Moreover, though grouping queries are possible in XQuery, there are many issues with the results \cite{beyer05}. We nonetheless select XQuery for expressing XWeB's workload due to its standard status. Furthermore, introducing difficult queries in the workload aims to challenge XML DBMS query engines.

Although we do take inspiration from TPC-H and SSB, our particular XML warehouse schema leads us to propose yet another query workload. It is currently composed of twenty decision support queries labeled Q01 to Q20 that basically are typical aggregation queries for decision support. Though we aim to provide the best functional and selectivity coverages with this workload, we lack experimental feedback, thus it is likely to evolve in the future. Workload specification is provided in Table~\ref{Table:workload}. Queries are presented in natural language for space constraints, but their complete XQuery formulation is available on-line (Section~\ref{sec:Conclusion}). 

\begin{table*}[hbtp]
	 \caption{XWeB workload specification}
	 \label{Table:workload}
   \begin{center}
      \begin{tabular}{cclcc}
			\hline \noalign{\smallskip}
         \scriptsize{Group} & \scriptsize{Query} & \scriptsize{Specification} & \scriptsize{Restriction} & \scriptsize{Ordering} \\
         \noalign{\smallskip} \hline \noalign{\smallskip}
         \scriptsize{Reporting} & \scriptsize{Q01} & \scriptsize{Min, Max, Sum, Avg of}  & & \\
         & & \scriptsize{$f\_quantity$ and $f\_totalamount$} & & \\
         \hline
         & \scriptsize{Q02} & \scriptsize{$f\_quantity$ for each $p\_partkey$} & \scriptsize{$p\_retailprice \leq 1000$} & \scriptsize{$p\_retailprice$} \\
         \hline
         & \scriptsize{Q03} & \scriptsize{Sum of $f\_totalamount$} & \scriptsize{$n\_name = "FRANCE"$} &  \\
         \hline 
         \scriptsize{1D cube} & \scriptsize{Q04} & \scriptsize{Sum of $f\_quantity$ per $p\_partkey$} & \scriptsize{$p\_retailprice > 1500$} & \scriptsize{$p\_retailprice^{-1}$} \\
         \hline
         & \scriptsize{Q05} & \scriptsize{Sum of $f\_quantity$ and $f\_total$-} & \scriptsize{Quarter($m\_monthkey) = 1$} & \scriptsize{$m\_monthname$} \\
         & & \scriptsize{$amount$ per $m\_monthname$} & & \\
         \hline
         & \scriptsize{Q06} & \scriptsize{Sum of $f\_quantity$ and $f\_total$-} & \scriptsize{Quarter($m\_monthkey) = 1$} & \scriptsize{$d\_dayname$} \\
         & & \scriptsize{$amount$ per $d\_dayname$} & & \\
         \hline
         & \scriptsize{Q07} & \scriptsize{Avg of $f\_quantity$ and $f\_total$-} & \scriptsize{$r\_name = "AMERICA"$} &  \\
         & & \scriptsize{$amount$ per $r\_name$} & & \\
         \hline
         \scriptsize{2D cube} & \scriptsize{Q08} & \scriptsize{Sum of $f\_quantity$ and $f\_total$-} & \scriptsize{$p\_brand = "Brand\#25"$} & \scriptsize{$c\_name,$} \\
         & & \scriptsize{$amount$ per $c\_name$ and $p\_name$} & & \scriptsize{$p\_name$} \\         
         \hline
         & \scriptsize{Q09} & \scriptsize{Sum of $f\_quantity$ and $f\_total$-} & \scriptsize{$p\_brand = "Brand\#25"$} & \scriptsize{$n\_name,$} \\
         & & \scriptsize{$amount$ per $n\_name$ and $p\_name$} & & \scriptsize{$p\_name$} \\     
         \hline
         & \scriptsize{Q10} & \scriptsize{Sum of $f\_quantity$ and $f\_total$-} & \scriptsize{$p\_brand = "Brand\#25"$} & \scriptsize{$r\_name,$} \\
         & & \scriptsize{$amount$ per $r\_name$ and $p\_name$} & & \scriptsize{$p\_name$} \\   
			\hline
         & \scriptsize{Q11} & \scriptsize{Max of $f\_quantity$ and $f\_total$-} & \scriptsize{$s\_acctbal < 0$} & \scriptsize{$s\_name,$} \\
         & & \scriptsize{$amount$ per $s\_name$ and $p\_name$} & & \scriptsize{$p\_name$} \\           
         \hline
         \scriptsize{3D cube} & \scriptsize{Q12} & \scriptsize{Sum of $f\_quantity$ and $f\_total$-} &  & \scriptsize{$c\_name,$} \\
         & & \scriptsize{$amount$ per $c\_name$, $p\_name$} & & \scriptsize{$p\_name,$} \\         
         & & \scriptsize{and $y\_yearkey$} & & \scriptsize{$y\_yearkey$} \\ 
         \hline
         & \scriptsize{Q13} & \scriptsize{Sum of $f\_quantity$ and $f\_total$-} & \scriptsize{$y\_yearkey > 2000$} & \scriptsize{$c\_name,$} \\
         & & \scriptsize{$amount$ per $c\_name$, $p\_name$} & \scriptsize{and $c\_acctbal > 5000$} & \scriptsize{$p\_name,$} \\         
         & & \scriptsize{and $y\_yearkey$} & & \scriptsize{$y\_yearkey$} \\ 
         \hline      
         & \scriptsize{Q14} & \scriptsize{Sum of $f\_quantity$ and $f\_total$-} & \scriptsize{$c\_mktsegment = "AUTO$-} & \scriptsize{$c\_name,$} \\
         & & \scriptsize{$amount$ per $c\_name$, $p\_name$} & \scriptsize{$MOBILE"$} & \scriptsize{$p\_name,$} \\         
         & & \scriptsize{and $y\_yearkey$} & \scriptsize{and $y\_yearkey = 2002$} & \scriptsize{$y\_yearkey$} \\ 
         \hline
         \scriptsize{Complex} & \scriptsize{Q15} & \scriptsize{Avg of $f\_quantity$ and $f\_total$-} &  & \scriptsize{$t\_name$} \\
         \scriptsize{hierarchy} & & \scriptsize{$amount$ per $t\_name$} & & \\  
         \hline
         & \scriptsize{Q16} & \scriptsize{Avg of $f\_quantity$ and $f\_total$-} & \scriptsize{$t\_name = "BRUSHED"$} & \scriptsize{$t\_name$} \\
         & & \scriptsize{$amount$ per $t\_name$} & & \\   
         \hline
         & \scriptsize{Q17} & \scriptsize{Avg of $f\_quantity$ and $f\_total$-} & \scriptsize{$t\_name = "BRUSHED"$} & \scriptsize{$p\_name$} \\
         & & \scriptsize{$amount$ per $p\_name$} & & \\  
         \hline
         & \scriptsize{Q18} & \scriptsize{Sum of $f\_quantity$ and $f\_total$-} & \scriptsize{$p\_size > 40$} & \scriptsize{$p\_name$} \\
         & & \scriptsize{$amount$ per $p\_name$} & & \\ 
         \hline
         & \scriptsize{Q19} & \scriptsize{Sum of $f\_quantity$ and $f\_total$-} & \scriptsize{$p\_size > 40$} & \scriptsize{$t\_name$} \\
         & & \scriptsize{$amount$ per $t\_name$} & & \\   
         \hline
         & \scriptsize{Q20} & \scriptsize{Sum of $f\_quantity$ and $f\_total$-} & \scriptsize{$p\_size > 40$} & \scriptsize{$t\_name$} \\
         & & \scriptsize{$amount$ per $t\_name$} &  & \\                                                               
         \hline
      \end{tabular}
   \end{center}
\end{table*}

XWeB's workload is roughly structured in increasing order of query complexity, starting with simple aggregation, then introducing join operations, then OLAP-like queries such as near-cube (with superaggregates missing) calculation, drill-downs (e.g., Q06 drills from Q05's \emph{Month} down to \emph{Day} granularity) and rollups (e.g., Q09 rolls from Q08's \emph{Customer} up to \emph{Nation} granularity), while increasing the number of dimensions involved. The last queries exploit the \emph{Part/Category} complex hierarchy. We also vary the type of restrictions (by-value and range queries), the aggregation function used, and the ordering applied to queries. Ordering labeled by $^{-1}$ indicates a descending order (default being ascending). Finally, note that Q20, though apparently identical to Q19, is a further rollup along the \emph{Category} complex hierarchy. Actually, Q19 rolls up from Q18's product level to the category level, and then Q20 rolls up to the ``supercategory'' level, with supercategories being categories themselves.

Moreover, workload queries are subdivided into five categories: simple reporting (i.e., non-grouping) queries; 1, 2, and 3-dimension cubes; and complex hierarchy cubes. We indeed notice in our experiments (Section~\ref{sec:Experiments}) that complex queries are diversely handled by XML DBMSs: some systems have very long response times, and even cannot answer. Subdividing the workload into blocks allows us to adjust workload complexity, by introducing boolean execution parameters ($RE$, $1D$, $2D$, $3D$ and $CH$, respectively) that define whether a particular block of queries must be executed or not when running the benchmark (see below).

\subsubsection{Execution Protocol and Performance Metrics.}
\label{sec:Execution}

Still with TPC-H as a model, we adapt its execution protocol along two axes. First, since XWeB does not currently feature update operations (Section~\ref{sec:Principle}), the performance test can be simplified to executing the query workload. Second, as in DWEB, we allow warm runs to be performed several times (parameter $NRUN$) instead of just once, to allow averaging results and flattening the effects of any unexpected outside event. Thus, the overall execution protocol may be summarized as follows:
\begin{enumerate}
	\item \textbf{load test:} load the XML warehouse into an XML DBMS; 
	\item \textbf{performance test:}
		\begin{enumerate}
			\item \emph{cold run} executed once (to fill in buffers), w.r.t. parameters $RE$, $1D$, $2D$, $3D$ and $CH$;
			\item \emph{warm run} executed $NRUN$ times, still w.r.t. workload parameters.
		\end{enumerate}
\end{enumerate}

The only performance metric in XWeB is currently \emph{response time}, as in SSB and DWEB. Load test, cold run and warm runs are timed separately. Global, average, minimum and maximum execution times are also computed, as well standard deviation. This kind of atomic approach for assessing performance allows to derive any more complex, composite metrics, such as TPC-H's throughput and power if necessary, while remaining simple.

\section{Sample Experiments}
\label{sec:Experiments}

To illustrate XWeB's usage, we compare in this section a sample of XML-native DBMSs, namely BaseX\footnote{http://www.inf.uni-konstanz.de/dbis/basex/}, eXist\footnote{http://exist.sourceforge.net}, Sedna\footnote{http://www.modis.ispras.ru/sedna/}, X-Hive\footnote{http://www.emc.com/domains/x-hive/} and xIndice\footnote{http://xml.apache.org/xindice/}. We focus on XML-native systems in these experiments because they support the formulation of decision support XQueries that include join operations, which are much more difficult to achieve in XML-compatible relational DBMSs. In these systems, XML documents are indeed customarily stored in table rows, and XQueries are embedded in SQL statements that target one row/document, making joins between XML documents difficult to express and inefficient.

Our experiments atomize the execution protocol from Section~\ref{sec:Execution}, on one hand to separately outline how its steps perform individually and, on the other hand, to highlight performance differences among the studied systems. Moreover, we vary data warehouse size (expressed in number of facts) in these experiments, to show how the studied systems scale up. Table~\ref{Table:size-documents} provides the correspondence between the number of facts, parameters $SF$ and $D$, and warehouse size in kilobytes. Note that warehouse sizes are small because most of the studied systems \emph{do not} scale up on the hardware configuration we use (a Pentium 2~GHz PC with 1~GB of main memory and an IDE hard drive running under Windows~XP). The possibility of missing values and element reordering is also disregarded in these preliminary experiments, i.e., $P_m = P_o = 0$. 

\begin{table*}[hbt]
   \caption{Total size of XML documents}
   \label{Table:size-documents}
   \begin{center}
      \begin{tabular}{ccrr}
			\hline \noalign{\smallskip} $SF$ & $D$ & Number of facts & Warehouse size (KB) \\
      \noalign{\smallskip} \hline \noalign{\smallskip} 1 & $1/14 \times 10^{-7}$ & 500 &  1710 \\
      \hline \noalign{\smallskip} 1 & $1/7 \times 10^{-7}$ & 1000 & 1865\\
      \hline \noalign{\smallskip} 1 & $2/7 \times 10^{-7}$ & 2000 & 2139\\
			\hline \noalign{\smallskip} 1 & $3/7 \times 10^{-7}$ & 3000 & 2340\\          			
      \hline \noalign{\smallskip} 1 & $4/7 \times 10^{-7}$ & 4000 & 2686\\          			
      \hline \noalign{\smallskip} 1 & $5/7 \times 10^{-7}$ & 5000 & 2942\\
      \hline \noalign{\smallskip} 1 & $6/7 \times 10^{-7}$ & 6000 & 3178\\          			
      \hline \noalign{\smallskip} 1 & $10^{-7}$ & 7000 & 3448\\          			
      \hline
      \end{tabular}
   \end{center}
\end{table*}

\subsection{Load Test}
\label{subsec:load}

\begin{figure}[hbt]
{\centering
\resizebox*{0.7\textwidth}{!}{\includegraphics{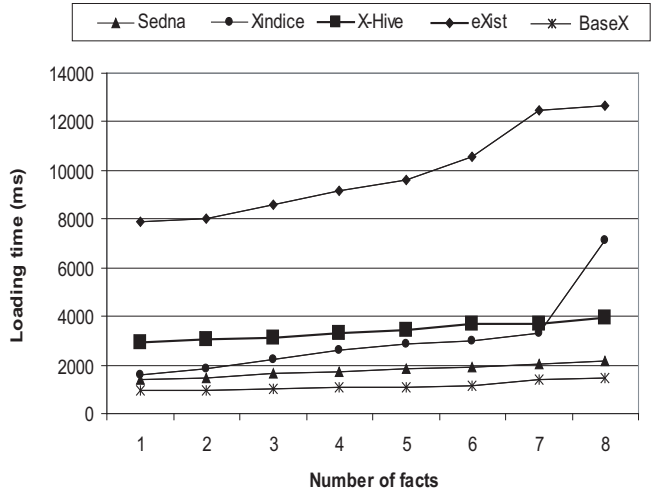}}
\par}
\caption{Load test results} 
\label{fig:load-time}
\end{figure}

Figure~\ref{fig:load-time} represents loading time with respect to data warehouse size. We can cluster the studied systems in three classes. BaseX and Sedna feature the best loading times. BaseX is indeed specially designed for full-text storage and allows compact and high-performance database storage, while Sedna divides well-formed XML documents into parts of any convenient size before loading them into a database using specific statements of the Data Manipulation Language. Both these systems load data about twice faster than X-Hive and xIndice, which implement specific numbering schemes that optimize data access, but require more computation at storage time, especially when XML documents are bulky. Finally, eXist performs about twice worse than X-Hive and xIndice because, in addition to the computation of a numbering scheme, it builds document, element and attribute indexes at load time.

\subsection{Performance Test}
\label{subsec:workload}

In this set of experiments, we measure query execution time with respect to data warehouse size. Since we rapidly test the limits of the studied systems, we only and separately evaluate the response of reporting, 1-dimension cube, and complex hierarchy-based queries, respectively. In terms of workload parameters, $RE = 1D = CH = TRUE$ and $2D = 3D = FALSE$. Moreover, we stop time measurement when workload execution time exceeds three hours. Finally, since we perform atomic performance tests, they are only cold runs (i.e., $NRUN = 0$). 

\begin{figure}[hbt]
{\centering
\resizebox*{0.9\textwidth}{!}{\includegraphics{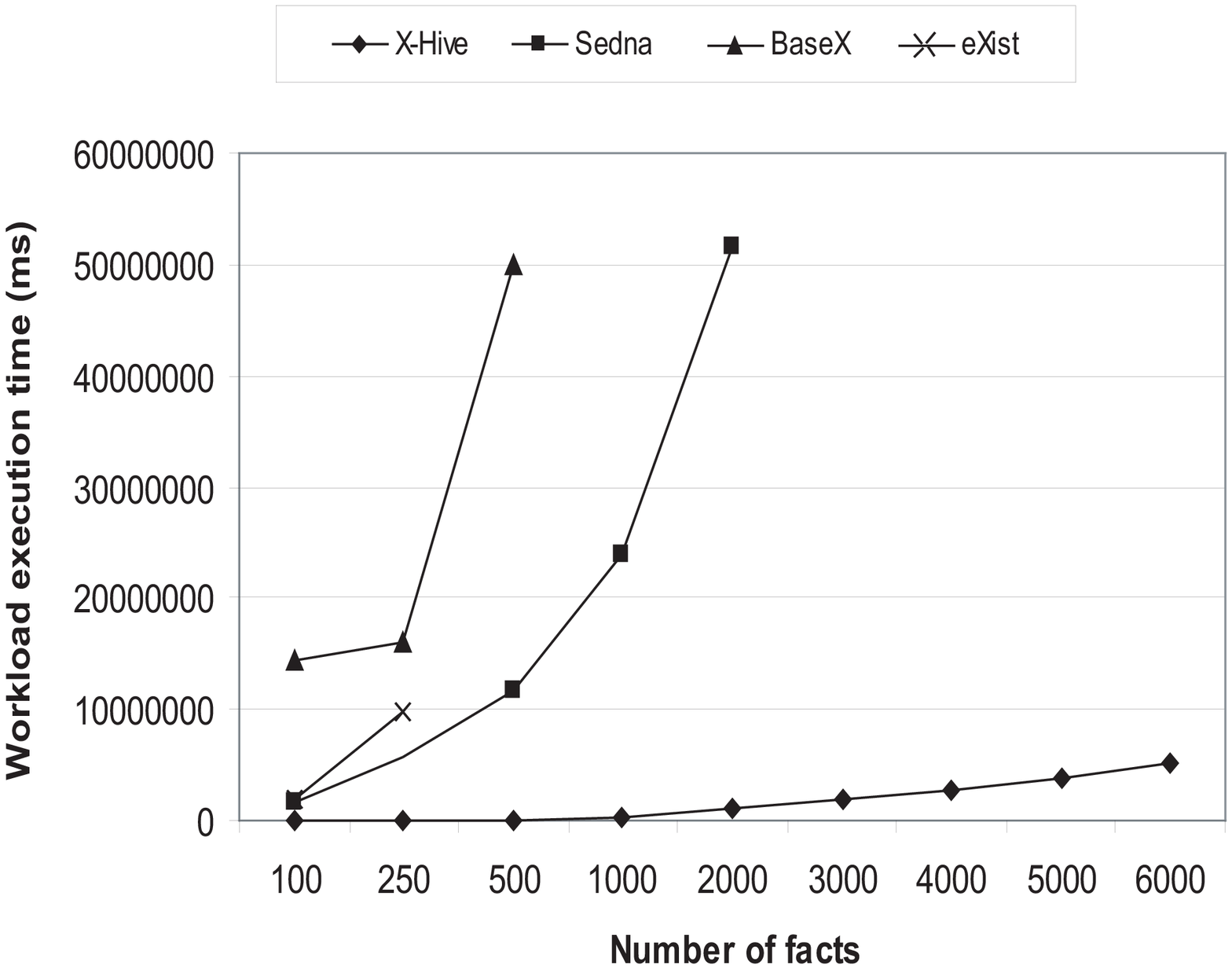}}
\par}
\caption{RE performance test results} 
\label{fig:reporting}
\end{figure}

Figure \ref{fig:reporting} represents the execution time of reporting queries (RE) with respect to warehouse size. Results clearly show that X-Hive's claimed scalability capability is effective, while the performance of other systems degrades sharply when warehouse size increases. We think this is due to X-Hive's specifically designed XProc query Engine (a pipeline engine), while Sedna and BaseX are specially designed for full-text search and do not implement efficient query engines for structural query processing. Finally, eXist and xIndice are specifically adapted to simple XPath queries processed on a single document and apparently do not suit complex querying needs.

\begin{figure}[hbt]
{\centering
\resizebox*{0.9\textwidth}{!}{\includegraphics{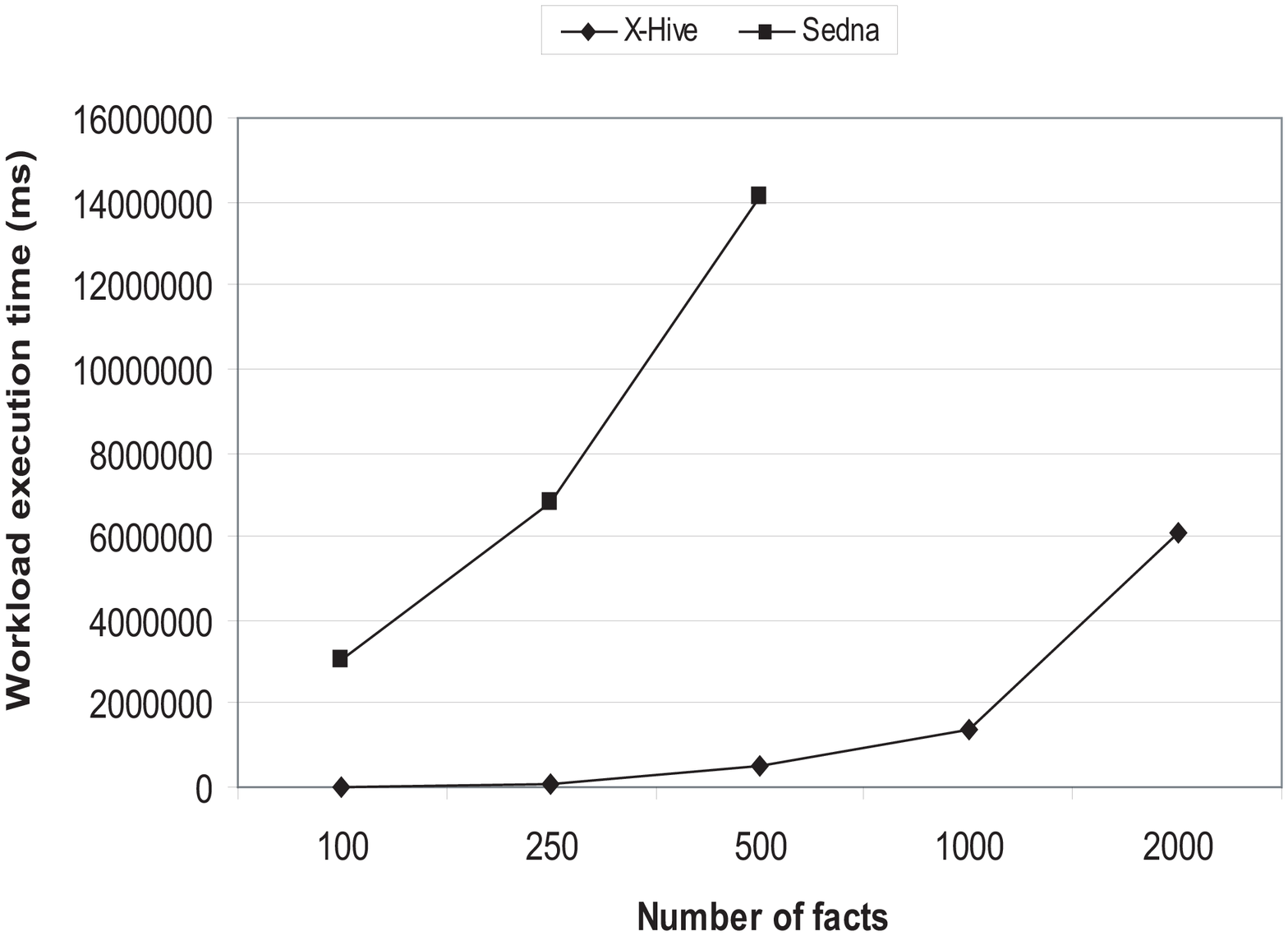}}
\par}
\caption{1D performance test results} 
\label{fig:1D}
\end{figure}

In Figure \ref{fig:1D}, we plot the execution time of 1D cube queries (1D) with respect to warehouse size. We could only test Sedna and X-Hive here, the other systems being unable to execute this workload in a reasonable time (less than three hours). X-Hive appears the most robust system in this context. This is actually why we do not push toward the 2D and 3D performance tests. Only X-Hive is able to execute these queries. With other systems, execution time already exceeds three hours for one single query. The combination of join and grouping operations (which induce further joins in XQuery) that are typical in decision support queries should thus be the subject of dire optimizations. 

\begin{figure}[hbt]
{\centering
\resizebox*{0.7\textwidth}{!}{\includegraphics{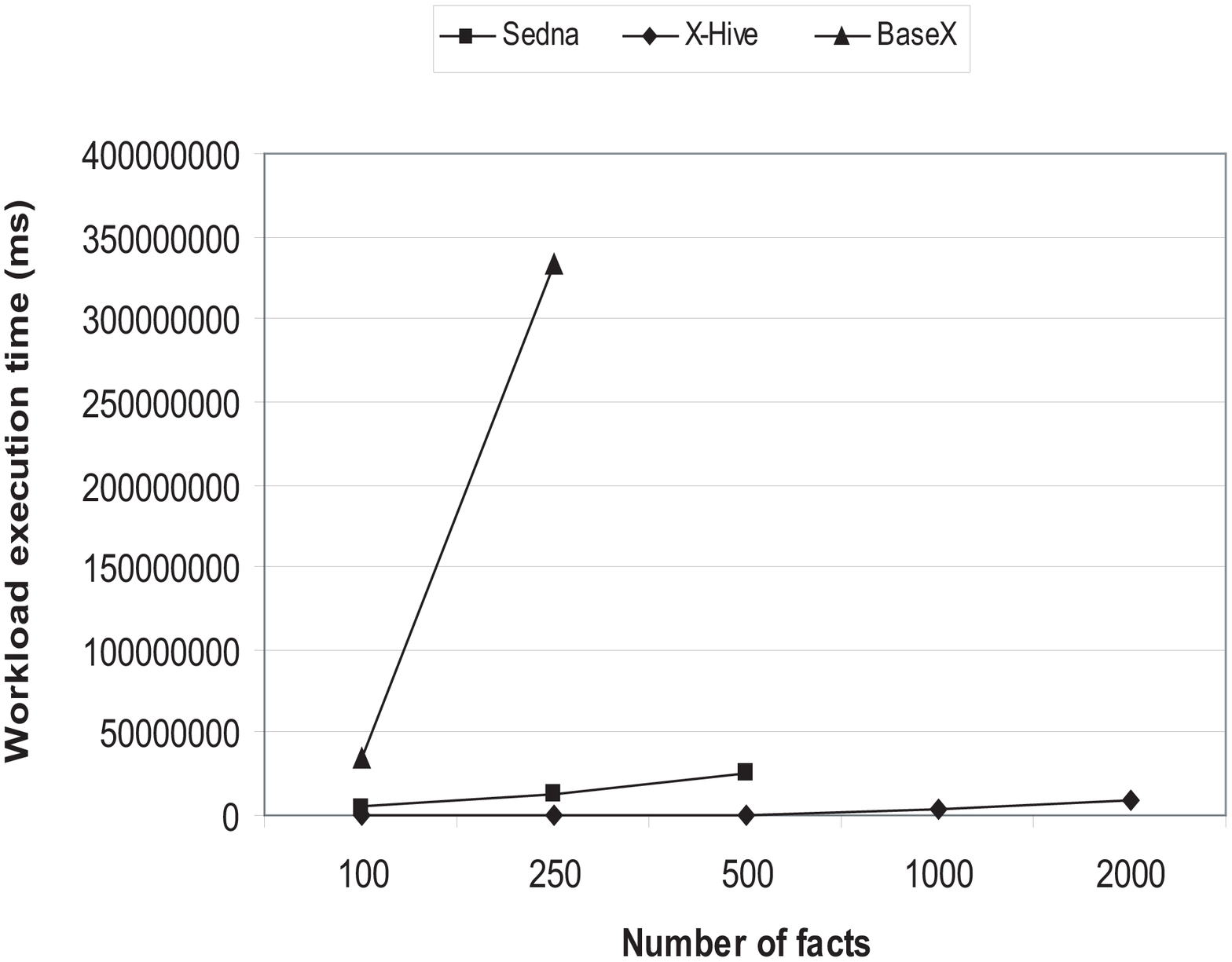}}
\par}
\caption{CH performance test results} 
\label{fig:complex}
\end{figure}

Finally, Figure \ref{fig:complex} features the execution time of complex hierarchy-based queries (CH) with respect to warehouse size. In this test, we obtained results only with X-Hive, Sedna and BaseX. Again, X-Hive seems the only XML-native DBMS to be able to scale up with respect to warehouse size when multiple join operations must be performed.

\section{Conclusion and Perspectives}
\label{sec:Conclusion}

When designing XWeB, which is to the best of our knowledge the first XML decision support benchmark, we aimed at meeting the four key criteria that make a ``good'' benchmark according to Jim Gray~\cite{GRA93}. \emph{Relevance} means the benchmark must answer various engineering needs. This is why we chose to base our work on a TPC standard. We also introduced more tunability, both at schema and workload levels, to adapt to the reality of XML DBMSs. \emph{Portability} means the benchmark must be easy to implement on different systems. To this aim, we implemented XWeB with the Java language that allows connecting to most XML DBMSs through APIs (we used the very popular XML:DB\footnote{http://xmldb-org.sourceforge.net/xapi/}). \emph{Scalability} means it must be possible to benchmark small and large databases, and to scale up the benchmark, which is achieved by inheriting from the $SF$ parameter. Further tuning is achieved through the density ($D$) parameter. Eventually, \emph{simplicity} means that the benchmark must be understandable, otherwise it will not be credible nor used. This is why we elected to base XWeB on TPC-H rather than TPC-DS or DWEB.

In this paper, we also illustrated XWeB's relevance through several experiments aimed at comparing the performance of five native-XML DBMSs. Although basic and more focused on demonstrating XWeB's features than comparing the studied systems in depth, they highlight X-Hive as the most scalable system, while full-text systems such as BaseX seem to feature the best data storage mechanisms. Due to equipment limitations, we remain at small scale factors, but we believe our approach can be easily followed for larger scale factors. We also show the kind of decision support queries that require urgent optimization: namely, cubing queries that perform join and grouping operations on a fact document and dimension documents. In this respect, XWeB had previously been successfully used to experimentally validate indexing and view materialization strategies for XML data warehouses \cite{asd06}.

Eventually, a raw, preliminary version of XWeB (warehouse, workload, Java interface and source code) is freely available online\footnote{http://ena-dc.univ-lyon2.fr/download/xweb.zip} as an Eclipse\footnote{http://www.eclipse.org} project. A more streamlined version is in the pipe and will be distributed under Creative Commons licence\footnote{http://creativecommons.org/licenses/by-nc-sa/2.5/}.

After having designed a benchmark modeling business \emph{data} (which XWeB aims to be), it would be very interesting in future research to also take into account the invaluable business information that is stored into unstructured documents. Hence, including features from, e.g., XBench into XWeB would help improve a decision support benchmark's XML specificity.

Since the XQuery Update Facility has been issued as a candidate recommendation by the W3C \cite{xqu09} and is now implemented in many XML DBMSs (e.g., eXist, BaseX, xDB, DB2/PureXML, Oracle Berkeley DB XML...), it will also be important to include update operations in our workload. The objective is not necessarily to feature full ETL testing capability, which would presumably necessitate a dedicated benchmark (Section~\ref{sec:Principle}), but to improve workload relevance with refreshing operations that are casual in data warehouses, in order to challenge system response and management of redundant performance optimization structures such as indexes and materialized views.

The core XWeB workload (i.e., read accesses) shall also be given attention. It has indeed been primarily designed to test scaling up. Filter factor analysis of queries \cite{oneil09} and experimental feedback should help tune it and broaden its scope and representativity. Moreover, we mainly focus on cube-like aggregation queries in this version. Working on the output cubes from these queries might also be interesting, i.e., by applying other usual XOLAP operators such as slice~\&~dice or rotate that are easy to achieve in XQuery \cite{datax08}.

Finally, other performance metrics should complement response time. Beyond composite metrics such as TPC benchmarks', we should not only test system response, but also the quality of results. As we underlined in Section~\ref{sec:WorkloadQueries}, complex grouping XQueries may return false answers. Hence, query result correctness or overall correctness rate could be qualitative metrics. Since several XQuery extension proposals do already support grouping queries and OLAP operators \cite{beyer05,datax08,KitAK09,WiwatwattanaJLS07}, we definitely should be able to test systems in this regard.

\bibliographystyle{splncs03}
\bibliography{xweb}  

\end{document}